\documentclass[doublecol]{epl2} 

\usepackage{graphicx,epsfig,amsmath}
\newcommand {\mcu}{\mathcal{U}}

\title {Shallow Efimov tetramer as inelastic virtual state
and resonant enhancement of the atom-trimer relaxation}
\shorttitle{Shallow Efimov tetramer} 

\author{A. Deltuva} 
\institute{
Centro de F\'{\i}sica Nuclear da Universidade de Lisboa, P-1649-003 Lisboa, Portugal }


\abstract
{We use exact four-boson scattering equations in the momentum-space
framework to study the universal properties of shallow Efimov tetramers
and their dependence on  the two-boson scattering length. 
We demonstrate that, in contrast to previous predictions,
the shallow tetramer in a particular experimentally unexplored
regime is not an unstable bound state
but an inelastic virtual state. This leads to a resonant behaviour 
of the atom-trimer scattering length and thereby to
a resonant enhancement of the trimer relaxation in ultracold atom-trimer 
mixtures.}

\begin{document} 

 \maketitle

\section{Introduction}

The system of three identical bosons with short-range interactions
in the unitary limit, i.e.,
when the two-particle scattering length $a$ is infinite
and the dimer binding energy $b_d$ vanishes,
was first studied by V. Efimov in 1970 \cite{efimov:plb}.
He predicted  an infinite number of zero spin and positive parity ($0^+$)
trimer states with geometric spectrum and accumulation point
at the three-particle threshold, i.e.,
$b_{n}/b_{n-1} \approx 1/515$ for $n \to \infty$, 
where $b_n$ is the binding energy of the $n$th Efimov trimer. 
 {The reason for this phenomenon, called the Efimov effect,
 is that a resonant  two-particle 
interaction (large $|a|$) yields an effective long-range attraction
in the systems with $N \ge 3$ particles  \cite{efimov:plb}. 
The resulting few-body bound states (except maybe for few lowest of them) 
are of non-clasical nature, i.e., their size is much larger than the range 
of the two-body interaction with the particles  residing predominantly
in an interaction-free region. As a consequence, the properties of such
few-body systems are universal, i.e.,  independent of the details of the 
short-range interaction \cite{braaten:rev}. A characteristic feature 
in bosonic systems is a log-periodic $a$-dependence of  few-body observables
 \cite{braaten:rev}.

The experimental evidence for the  Efimov physics
 was observed 35 years later  in  systems of ultracold atoms
where the formation of the Efimov trimers led to a resonant enhancement or 
suppression of the recombination or relaxation processes
\cite{kraemer:06a,knoop:09a,PhysRevLett.103.043201};
large values of $a$ were 
achieved by variation of the external magnetic field in the 
vicinity of the Feshbach resonance. First steps in exploring the
four-atom Efimov physics via ultracold atom experiments 
\cite{ferlaino:09a,pollack:09a} have been made recently calling
for accurate theoretical studies of the four-body systems with large $|a|$.}

\begin{figure}[!]
\includegraphics[scale=0.45]{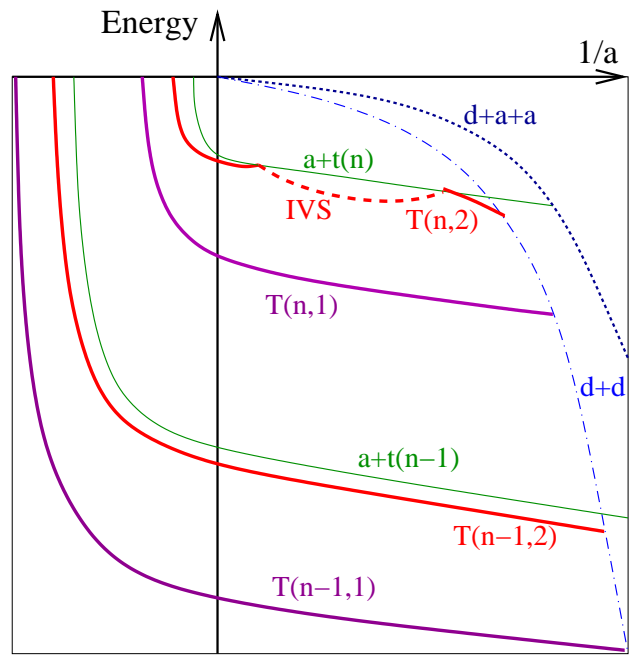} 
\caption{\label{fig:efi} 
 {
Schematic representation of the four-boson energy spectrum
as a function of the two-boson scattering length.
The tetramer (T) energies, atom-trimer (a+t),  dimer-dimer (d+d),
and dimer-atom-atom (d+a+a) thresholds are shown as  
thick solid, thin solid, dashed-dotted, and dotted curves,
respectively. For a better visualization only qualitative but not
quantitative relations between them are preserved and 
only two families of multimers, the $(n-1)$th and $n$th, are shown. 
Positive (negative) $a$ are on the right (left) side from the
vertical axis while the four free atom threshold 
lies on the upper horizontal axis.
The behaviour of the $(n-1,2)$th tetramer
corresponds to the predictions of  ref.~\cite{stecher:09a}
and is non-universal whereas that  of the  $(n,2)$th tetramer is universal;
the dashed part of the curve corresponds to the IVS. }}
\end{figure}

Although there is no Efimov effect in the four-boson system,
the three-boson Efimov effect has an impact on the 
atom-trimer scattering observables: scattering lengths,
phase shifts, elastic and inelastic cross sections are related to the
corresponding trimer binding energies in a universal way \cite{deltuva:10c}.
 { 
Furthermore,  it was predicted in refs.~\cite{hammer:07a,stecher:09a}
that below each Efimov trimer two tetramers with spin/parity $0^+$
should exist. We label them by two integers $(n,k)$ where $n$ refers
to the associated trimer and $k=1$ (2) for a more deeply 
(shallowly) bound tetramer. A schematic  representation of the four-boson 
energy levels as functions of the two-boson scattering length
$a$ is given in fig.~\ref{fig:efi}.
 For a better visualization only  two families of multimers
are shown preserving only qualitative  relations between them.}
The two lowest  tetramers with $n=0$ are true bound states;
all others lie above the lowest atom-trimer threshold (ATT)
and therefore are unstable bound states (UBS) 
with finite width and lifetime that could not be calculated
in refs.~\cite{hammer:07a,stecher:09a} using standard bound state techniques. 
The widths of unstable tetramers in the unitary limit
were determined using proper four-boson
scattering calculations \cite{deltuva:10c}.
Trimers and the associated  tetramers exist also at large finite $|a|$ 
as predicted in refs.~\cite{hammer:07a,stecher:09a,dincao:09a} and  
confirmed in experiments with ultracold atoms \cite{ferlaino:09a,pollack:09a}. 
At specific large negative values of $a$ where the tetramers emerge
at the four free atom threshold they lead to a
resonant enhancement of the four-atom recombination process
\cite{stecher:09a,ferlaino:09a}.
At specific large positive  values of $a$ where the tetramers connect to
the dimer-dimer threshold they lead to a resonant increase of
the dimer-dimer relaxation rate \cite{dincao:09a}.
It was predicted in ref.~\cite{stecher:09a}
that the two tetramers do not intersect the associated ATT
 and lie below it in the whole region of their existence
between the four-atom and the dimer-dimer thresholds.
 { This scenario is schematically shown in fig.~\ref{fig:efi}
for the $(n-1)$th (lower) multimer family.}
However, in the present work we prove that this is not true
for the shallow tetramer in a particular experimentally not yet explored 
interval of positive $a$ values where it crosses the ATT and 
becomes an inelastic virtual state (IVS);
 { this scenario is schematically shown in fig.~\ref{fig:efi}
for the $n$th (higher) multimer family.}  As a consequence,
 the atom-trimer scattering length shows a resonant behaviour that would
lead to a resonant enhancement of the trimer relaxation in an ultracold 
mixture of atoms and excited Efimov trimers.
Our proof relies on rigorous four-boson scattering calculations
in the momentum-space framework. With some technical modifications
\cite{deltuva:10c} the method follows the
one of refs.~\cite{deltuva:07a,deltuva:07b,deltuva:07c}
that has already been applied successfully to the description of
all four-nucleon elastic and transfer reactions at low energies.

\section{Atom-trimer scattering}

We use exact four-particle scattering equations that were introduced by 
Alt, Grassberger, and Sandhas (AGS) \cite{grassberger:67,alt:jinr};
they are  equivalent to the Faddeev-Yakubovsky formulation 
\cite{yakubovsky:67en} of the four-body problem.
The AGS equations are integral equations for the  four-particle
transition operators $\mcu_{\beta\alpha}$; in the case of identical particles
they are given in refs.~\cite{deltuva:10c,deltuva:07a}.
In the present work we are interested in the energy regime where 
only the atom-trimer channels are open and 
the  AGS equations can formally be reduced to a single integral equation
\begin{eqnarray} \nonumber
\mcu_{11}  &=&  P_{34} (G_0  t  G_0)^{-1}  +  U_2(1 + P_{34})
 + P_{34}  U_1 G_0  t G_0  \mcu_{11} \\ && +
U_2 G_0  t G_0  (1 + P_{34}) U_1 G_0  t  G_0  \mcu_{11}. \label{eq:U}
\end{eqnarray}
The two-boson potential $v$ enters the AGS equations via the
  two-body transition matrix $t = v + v G_0 t$
and  the symmetrized 1+3 ($\alpha=1$) and 
2+2 ($\alpha=2$) subsystem  transition operators
$ U_{\alpha} =  P_\alpha G_0^{-1} + P_\alpha  t G_0  U_{\alpha}$.
All transition operators acquire their dependence on the 
available energy $E$ via the free resolvent
$G_0 = (E+i0-H_0)^{-1}$ with $H_0$ being the free Hamiltonian.
 {
The  permutation operators $P_{ij}$ of particles $(ij)$ and 
their combinations $P_1 =  P_{12}\, P_{23} + P_{13}\, P_{23}$ and
$P_2 =  P_{13}\, P_{24} $ 
ensure the desired symmetry of the four-boson system.
Thus, in a simplified interpretation the integral equation (\ref{eq:U}) for the
transition operators $\mcu_{\beta\alpha}$ can be viewed as a sum of the
multiple scattering series with all possible interactions in the four-particle
system up to infinite order. Introducing the transition operators
$t$ and $ U_{\alpha}$ at intermediate steps ensures that
eq.~(\ref{eq:U}) is mathematically well behaved and can be solved 
numerically}. Amplitudes for elastic and inelastic atom-trimer scattering
$\langle f| T |i \rangle  = 3
\langle  \phi_{1}^f | \mcu_{11}| \phi_{1}^{i} \rangle$
are given by the on-shell matrix elements of the transition 
operator (\ref{eq:U}) between the Faddeev components
$ | \phi_1^{n} \rangle $ 
of the corresponding initial/final atom-trimer  states.

We use the momentum-space partial-wave framework 
\cite{deltuva:07a,deltuva:ef}
to solve the AGS equation (\ref{eq:U}) numerically. The practical solution
can be simplified significantly \cite{deltuva:10c} by
using an $S$-wave separable two-boson potential
$v = |g\rangle \lambda \langle g|$.
To prove that the universal properties of the four-boson system 
are independent of the short-range interaction details, 
as in ref.~\cite{deltuva:10c} we performed
calculations with two choices  of the momentum-space form factor
\begin{equation} \label{eq:gsep}
\langle p |g\rangle = [1+c_2\,(p/\Lambda)^2]e^{-(p/\Lambda)^2} ,
\end{equation}
namely, with $c_2 = 0$ and $c_2=-9.17$, that yield very different 
off-shell behaviour of the resulting potential whose
 strength $\lambda$ is adjusted to reproduce the given value of  $a$.
Further technical details on our four-boson scattering 
calculations can be found in 
refs.~\cite{deltuva:10c,deltuva:07a,deltuva:ef}.
In the case of the true bound state ($n=0$) we solve  Faddeev-Yakubovsky 
equations \cite{yakubovsky:67en}
in the version of ref.~\cite{deltuva:08a}.

\section{Extraction of tetramer properties}

For the classification of the states we follow ref.~\cite{res_cpl}.
A true four-boson bound state corresponds to a simple pole
of the transition operators  $\mcu_{\beta\alpha}$ in the 
physical sheet of the complex energy plane.
In contrast, an UBS corresponds to the pole of 
$\mcu_{\beta\alpha}$  in one of the unphysical sheets
 adjacent to the physical sheet  and therefore may lead to a 
resonance-like behaviour of the scattering observables. 
Let $E_{n,k} = -B_{n,k} - i\Gamma_{n,k}/2$ be the complex energy
of the $(n,k)$th unstable tetramer, where 
 $-B_{n,k}$ is the position relative to the four free particle  threshold 
and  $\Gamma_{n,k}$ is the width of the state.
The energy dependence of  $\mcu_{\beta\alpha}$ in
the physical region close to $E \approx -B_{n,k}$  can be given by
\begin{equation} \label{eq:Upole}
\mcu_{\beta\alpha} =  \sum_{m=-1}^\infty
\hat{\mcu}_{\beta\alpha}^{(n,k;m)} (E-E_{n,k})^{m}
\end{equation}
where only first few terms yield nonnegligible contributions.
This allows to extract the values of 
 $B_{n,k}$ and  $\Gamma_{n,k}$ from the calculated
atom-trimer scattering amplitudes or observables.
Since the shallow tetramer ($k=2$) is very close to the ATT,
$B_{n,2}$ and  $\Gamma_{n,2}$ alternatively
can be  obtained from the atom-trimer scattering length $A_n$ and 
effective range $r_n$ \cite{res_cpl,lazauskas:he}.
The elastic $S$-wave scattering amplitude 
as a function of the on-shell momentum $k_n$ is given by
\begin{equation} \label{eq:TS}
 T_n^S(k_n) = -[\pi \mu_1 (k_n \cot \delta_n^S - ik_n)]^{-1} 
\end{equation}
where $\mu_1$ is the reduced atom-trimer mass.
The complex phase shift  $\delta_n^S$ at small momenta $k_n$ can be
expanded as $k_n \cot \delta_n^S = -1/A_n + r_n k_n^2/2 + o(k_n^4)$.
Thus, the approximate value of the complex binding momentum $K_n$,
corresponding to the pole of the amplitude, can be obtained from
the equation 
\begin{equation} \label{eq:arK}
 r_n K_n^2/2 -iK_n -1/A_n = 0. 
\end{equation}
The root $K_n$ that is closer to the threshold, i.e.,
smaller in the absolute value,
relates to the tetramer position and width as
\begin{equation} \label{eq:KBG}
 -B_{n,2} - i\Gamma_{n,2}/2 = -b_n + K_n^2/2\mu_1 .
\end{equation}
For the UBS $\mathrm{Re} K_n < 0$,
$\mathrm{Im} K_n > 0$, and $\Gamma_{n,2} > 0$ \cite{res_cpl},
while for the true bound state  $\mathrm{Re} K_n = 0$,
$\mathrm{Im} K_n > 0$, and $\Gamma_{n,2} = 0$.
The method based on the effective range expansion (\ref{eq:arK}), 
although slightly less precise than
the one of eq.~(\ref{eq:Upole}), has an advantage of being applicable also
for the IVS in the $S$ wave that does not lead to a
resonant behaviour of the scattering observables around $E \approx -B_{n,2}$.
IVS corresponds to the pole of the amplitude (\ref{eq:TS}) 
with $\mathrm{Re} K_n < 0$, $\mathrm{Im} K_n < 0$ and $\Gamma_{n,2} < 0$ in the 
sheet of the complex energy plane more distant from the physical one
 \cite{res_cpl}. A particular case of the  $S$ wave IVS when
there is no lower threshold is
a virtual (unbound) state with $\mathrm{Re} K_n = 0$,
$\mathrm{Im} K_n < 0$, and $\Gamma_{n,2} = 0$.
In contrast, the   near-threshold unbound 
states with nonzero angular momentum  
have different properties and are called the Breit-Wigner resonances  
\cite{res_cpl}; a universal four-boson system doesn't support such
states \cite{deltuva:10c}.

\section{Results}

We present our results  in a universal form such
that they do not depend on the values of the particle mass 
or the cutoff parameter $\Lambda$. The binding energy of the
$n$th excited trimer $b_n$ and
the associated length scale $L_n = (2\mu_1 b_n)^{-1/2}$
(in our convention $\hbar=1$) will often be used to build dimensionless ratios, 
 { in particular, $B_{n,k}/b_n$  and $\Gamma_{n,2}/2b_n$ 
that are expected to be universal numbers.
To achieve the universal limit we have to consider
reactions with highly excited Efimov trimers that are of large
size such that the short-range details become negligible.}
This is a serious challenge in the numerical calculations
since the  binding energies of the involved trimers differ
by many orders of magnitude and therefore
the momentum grids of correspondingly broad range must be used. 
In this work we consider reactions with up to six open
atom-trimer channels and show that this is fully
sufficient to obtain universal results. 
The predictions for the positions and widths of shallow
tetramers in the unitary limit are collected in table~\ref{tab:BG2}.
The agreement between the two methods based on eqs.~(\ref{eq:Upole})
and  (\ref{eq:arK},\ref{eq:KBG}) is better than 0.5\% for $n>0$.
For highly excited tetramers, $n \ge 3$, the results approach 
the universal values for both choices of the form factor (\ref{eq:gsep}).
In contrast, large deviations can be seen for lower tetramers
where the finite range corrections become important.
\begin{table}[!]
\begin{tabular}{*{3}{c}} $n$ & 
 $(B_{n,2}/b_n-1) \times 10^{3}$  & $(\Gamma_{n,2}/2b_n) \times 10^{4}$ 
\\  \hline
0 &  41.9 &  \\
1 &  1.06 & 3.82 \\
2 &  2.17 & 2.14 \\
3 &  2.27 & 2.36 \\
4 &  2.28 & 2.38 \\
5 &  2.28 & 2.38 \\
\hline
1 &  9.97 & 4.18 \\
2 &  2.28 & 3.34 \\
3 &  2.27 & 2.39 \\
4 &  2.28 & 2.38 \\
\hline
\end{tabular}
\caption{ 
Positions and widths of shallow tetramers in the unitary limit.
Results obtained with $c_2=0$ ($c_2=-9.17$) in eq.~(\ref{eq:gsep})
are given in the top (bottom) part.}
\label{tab:BG2}
\end{table}

\begin{figure}[!]
\includegraphics[scale=0.66]{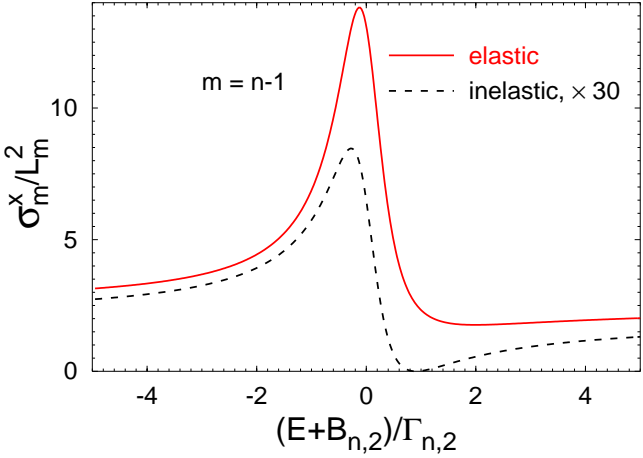} 
\caption{\label{fig:csu} 
Elastic and inelastic  cross sections 
for the atom-trimer scattering in the energy regime containing
the $(n,2)$th tetramer UBS. Unitary limit results 
for the highest available trimer ($m=n-1$) are shown.}
\end{figure}

In the following we do not demonstrate explicitly
the convergence of our results,
however, they are checked to be independent of $c_2$ and
$n$ for $n \ge 3$ with good accuracy. 
The energy dependence of the
elastic and inelastic cross sections for the atom scattering from
the highest available trimer ($m=n-1$) in the vicinity of the 
$(n,2)$th tetramer, i.e., at $E \approx -B_{n,2}$,
is shown in fig.~\ref{fig:csu}. Since the tetramer is UBS in the
unitary limit, both cross sections exhibit a resonant behaviour.
However, they deviate slightly from the exact  Breit-Wigner shape
due to non-negligible contributions of nonresonant (background) terms
corresponding to $m \ge 0$ in  eq.~(\ref{eq:Upole}).

\begin{figure}[!]
\includegraphics[scale=0.62]{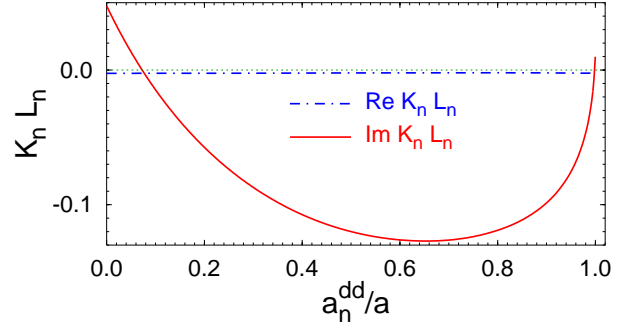} 
\caption{ 
Real and imaginary parts of the
complex binding momentum $K_n$ of the shallow tetramer
as functions of the two-boson scattering length $a$.}
\label{fig:K}
\end{figure}

\begin{figure}[!]
\includegraphics[scale=0.66]{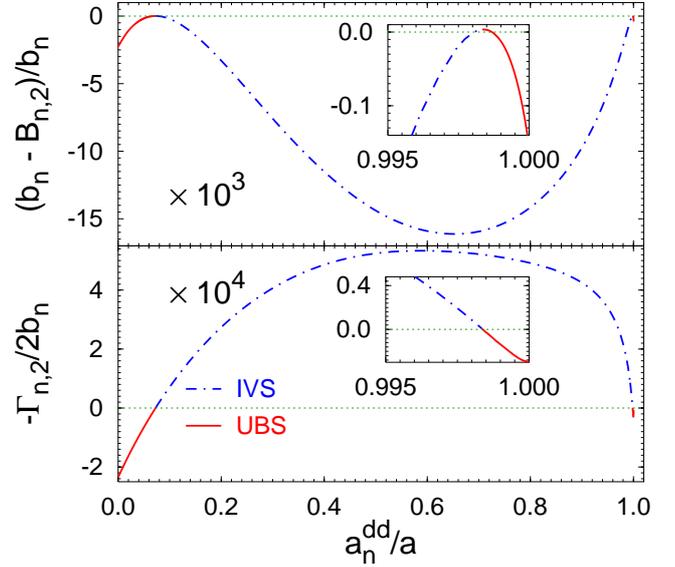} 
\caption{ 
Position of the shallow tetramer relative to the 
ATT (top) and its width (bottom)
as functions of the two-boson scattering length $a$.
 {The thin dotted line represents the zero value.}}
\label{fig:BG}
\end{figure}

We consider next the shallow tetramer in the
interval of positive two-boson scattering length $a$ between
the unitary limit $a \to \infty$
and the intersection of the atom-trimer and dimer-dimer 
thresholds $a = a_n^{dd}$ where $b_n = 2b_d$. 
The evolution of the binding momentum $K_n$ with $a$ is shown in
fig.~\ref{fig:K}; we use the dimensionless ratio $a_n^{dd}/a$.
 While $\mathrm{Re} K_n L_n$ remains negative and varies very slowly,
$\mathrm{Im} K_n$ changes the sign two times such that 
 $\mathrm{Im} K_n < 0$ for $a \in (a_n^{v,2},a_n^{v,1})$
and $\mathrm{Im} K_n > 0$ elsewhere. These critical values are
\begin{eqnarray} \label{eq:av1}
a_n^{dd}/a_n^{v,1} & \approx & 0.0729, \\ 
a_n^{dd}/a_n^{v,2} & \approx & 0.99836 .
\label{eq:av2}
\end{eqnarray}
Thus, the shallow tetramer is an
IVS for $a \in (a_n^{v,2},a_n^{v,1})$ but an UBS elsewhere.
The evolution of the tetramer properties is displayed in fig.~\ref{fig:BG}.
Solid parts of the curves correspond to UBS while the dash-dotted ones to IVS.
Starting at the unitary limit and increasing  $a_n^{dd}/a$ the tetramer
UBS moves towards the ATT while its width decreases.
$\Gamma_{n,2}$  passes through zero at $a = a_n^{v,1}$ where the tetramer
becomes an IVS. By further increase of $a_n^{dd}/a$ the tetramer IVS moves 
away from the ATT  but  turns around at
$a \approx 0.65 a_n^{dd}$ and  at $a = a_n^{v,2}$ where $\Gamma_{n,2} =0$ 
becomes an UBS again.
Note that in very narrow intervals close to $a = a_n^{v,j}$ the 
tetramer (in both UBS and IVS cases) is slightly
above the ATT, i.e., $B_{n,2} < b_n$.
This is evident from eq.~(\ref{eq:KBG}) with $\mathrm{Im} K_n = 0$
and can be seen in the topinset of fig.~\ref{fig:BG}.

We do not show the evolution of  the tetramer properties
for $a_n^{dd}/a > 1$ where our preliminary 
results are in qualitative agreement with ref.~\cite{dincao:09a}:
 the $(n,2)$th tetramer is an UBS with $B_{n,2} > 2b_d$ and
finite width until it intersects the dimer-dimer threshold at
 $a=a_{n,2}^{dd} < a_n^{dd}$.

\begin{figure}[!]
\includegraphics[scale=0.64]{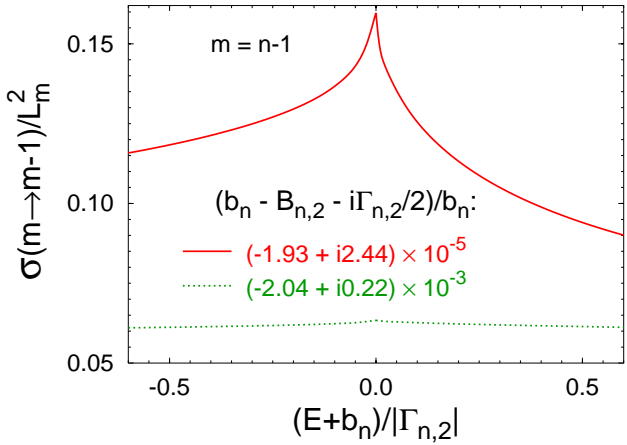} 
\caption{\label{fig:ivs} 
Cross section for the inelastic atom-trimer ($m=(n-1)$th state)
scattering leading to the ($m-1$)th trimer state
in the vicinity of the $n$th ATT. Two cases  differing in the
complex energy of the $(n,2)$th tetramer IVS are shown.}
\end{figure}

The tetramer IVS resides on a sheet of the complex 
energy plane that is not adjacent to the physical one.
It therefore affects the atom-trimer scattering observables in a completely
different way as compared to UBS: the cross sections around 
$E \approx -B_{n,2}$ do not exhibit the resonance-like behaviour 
seen in fig.~\ref{fig:csu} but may have a cusp at the ATT
($E = -b_n$) if the IVS pole is very close to it.
 This is illustrated in fig.~\ref{fig:ivs} where the atom-trimer
inelastic $(n-1 \to n-2)$ cross section is shown for two cases.
At $a_n^{dd}/a \approx 0.0820$ the tetramer IVS 
with $(b_n-B_{n,2}-i\Gamma_{n,2}/2)/b_n \approx (-1.93+i2.44) \times 10^{-5}$
is very close to the ATT  while it is more distant
at  $a_n^{dd}/a \approx 0.1682$ with 
$(b_n-B_{n,2}-i\Gamma_{n,2}/2)/b_n \approx (-2.04+i0.22) \times 10^{-3}$.
As a consequence, the cusp at the ATT is clearly
pronounced in the first case but hardly visible in the second.
We emphasize that such behaviour is characteristic to $S$ waves only
as the unbound states with nonzero angular momentum close to the threshold
are the Breit-Wigner resonances.

The resonant peak for the UBS and the cusp for the IVS can be understood
given the pole factor in the amplitude (\ref{eq:TS}) 
that is common for all elastic and inelastic channels. 
For simplicity we consider the case $|k_n|,\, |K_n| << 1/|r_n|$, i.e., when
the available scattering energy and the UBS/IVS poles are very close
to the $n$th ATT. Then the energy-dependence of the cross sections 
(up to channel-dependent factors) is approximately given by
\begin{eqnarray} \nonumber
\sigma  & \sim & | K_n - k_n|^{-2} \\  \nonumber
 & = &  
 [ (\mathrm{Re} K_n)^2 + (\mathrm{Im} K_n - \sqrt{2\mu_1|b_n+E|})^2]^{-1} \\ \nonumber
& \times &  \Theta(-b_n-E) \\ \nonumber
& + & [(\mathrm{Im} K_n)^2 + (\mathrm{Re} K_n - \sqrt{2\mu_1|b_n+E|})^2]^{-1} \\
& \times & \Theta(b_n+E). 
\label{eq:cs}
\end{eqnarray}
Here $\Theta(x)$ is the Heaviside step function with values 1 and 0 for $x>0$ and $x<0$,
respectively. Given  $\mathrm{Re} K_n < 0$ it is easy to see that
a resonant peak only takes place for $\mathrm{Im} K_n >0 $ at $E < -b_n$, i.e.,
for the UBS below the $n$th ATT. In contrast, for  $\mathrm{Im} K_n <0 $ 
the cross sections reach maximum value at $E = -b_n$ that corresponds to
the IVS cusp at the ATT.

\begin{figure}[!]
\includegraphics[scale=0.66]{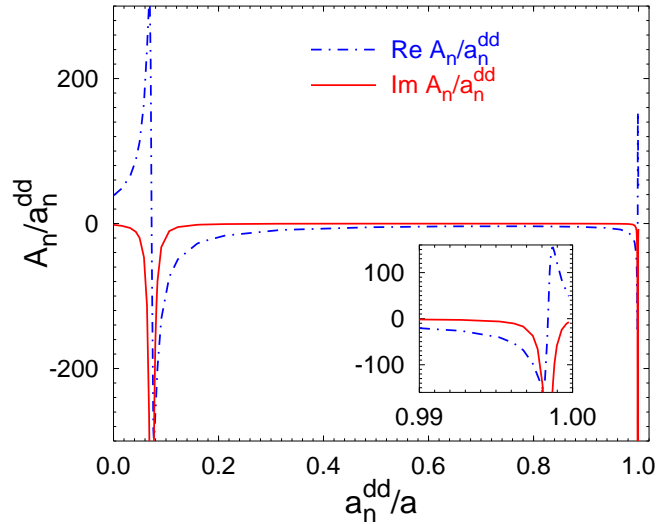}
\caption{Atom-trimer scattering length $A_n$
as function of the two-boson scattering length $a$.}
\label{fig:A}
\end{figure}

The two intersections of the ATT with the tetramer at $a = a_n^{v,j}$
lead to a resonant behaviour of the atom-trimer
scattering length $A_n$ as demonstrated in fig.~\ref{fig:A}.
The positive (negative) peaks of $\mathrm{Re} A_n$ correspond to
tetramer UBS (IVS) with $B_{n,2}=b_n$ while
$\mathrm{Re} A_n = 0$ and negative peaks of $\mathrm{Im} A_n$
correspond to $\Gamma_{n,2} = 0$.
No experiments have been performed so far in this regime but
the UBS-IVS conversion
 could be observed by creating an ultracold mixture 
of atoms and excited Efimov trimers in a trap and measuring
the trimer relaxation as function of $a$. 
According to ref.~\cite{deltuva:10c},
the zero-temperature limit of the trimer relaxation rate constant is
$\beta_n^0 = -(4\pi/\mu_1) \, \mathrm{Im} A_n$.
Thus, a resonant enhancement of the trimer relaxation process takes place
at $a = a_n^{v,j}$ that should be visible  at  sufficiently low
temperature, possibly  partially smeared out due to the thermal averaging.
On the other hand, in the real experiments deviations of $a_n^{v,j}$ 
from the universal values (\ref{eq:av1}-\ref{eq:av2}) can be expected
due to finite-range effects and/or presence of deeply bound dimers.

\section{Comparison with previous works}
Finally, we would like to discuss the differences between our results
and those of ref.~\cite{stecher:09a}. At the unitary limit
 ref.~\cite{stecher:09a} predicts $(B_{n,2}/b_n-1) \approx 0.01$ to be
more than 4 times larger than our universal result 0.00228
presented in table \ref{tab:BG2}. The results of ref.~\cite{stecher:09a}
were obtained using the  hyperspherical framework in  coordinate space
and neglecting the finite width of the tetramer. In addition, 
for the $(n,2)$th tetramer they were limited to $n \le 1$ ($n=0$
with strong repulsive three-body force) and by far have not reached 
the level of convergence comparable to ours.
 { For example, our five best-converged results for $(B_{n,2}/b_n-1)$
are 0.00228 within 0.5\% accuracy while
the three best-converged results of ref.~\cite{stecher:09a} are
0.006, 0.03, and 0.001. On the other hand, our  $n \le 1$
results given in table \ref{tab:BG2} 
that receive nonnegligible finite-range corrections
show deviations from the universal limit comparable to
those of  ref.~\cite{stecher:09a}.}
In the presence of finite-range effects we found 
that increasing  $(B_{n,2}/b_n-1)$  at $a \to \infty$
shifts  $a_n^{dd}/a_n^{v,1}$ to larger values thereby 
shrinking the IVS region. Thus,
when the tetramer  at $a \to \infty$ is sufficiently far  from the ATT,
it may even remain UBS (true bound state if $n=0$) 
in the whole regime $a_n^{dd}/a \le 1$. Such a situation takes place also
for the non-universal $(0,2)$th tetramer of the present work.  { 
However,  by choosing $c_2 > 1$ in eq.~(\ref{eq:gsep})
we obtain the $(0,2)$th tetramer that mimics the universal behaviour,
i.e., makes transition into  a virtual state. This proves again that 
low $n$ states are sensitive to the short-range details and may explain
why the tetramer IVS could not be predicted in ref.~\cite{stecher:09a}.

 In the case of the deeper ($k=1$) tetramer there is a
reasonanble agreement between our value  $B_{n,1}/b_n = 4.611(1)$ 
obtained in ref.~\cite{deltuva:10c} and 
the one of ref.~\cite{stecher:09a}, 4.58, 
the latter having about 1\% uncertainty.}

\section{Summary}

We studied the universal properties of the shallow Efimov tetramer.
Since this four-boson state lies in the continuum,
we solved exact AGS four-particle equations for the atom-trimer scattering.
The momentum-space partial-wave framework was employed.
The universal limit of the results is achieved accurately in 
 reactions with highly excited trimers (up to 5th excited state). 
Positions and widths of shallow tetramers were calculated
as functions of the two-boson scattering length. 
We demonstrated that, in contrast to previous predictions
by other authors \cite{stecher:09a,ferlaino:09a} obtained 
with some limitations, the  {universal} shallow tetramer intersects
the ATT twice and in a regime between those intersections  
 $a \in (a_n^{v,2},a_n^{v,1})$ it is an inelastic virtual state. 
The tetramer is an unstable bound state outside the above interval of $a$,
in qualitative agreement with the results of 
refs.~\cite{stecher:09a,dincao:09a}.
We studied how the atom-trimer scattering observables are affected by
these changes of the tetramer properties. In particular, we demonstrated
a resonant behaviour of the atom-trimer scattering length
that could be observed as a resonant enhancement of the trimer relaxation
in the ultracold mixture of atoms and excited trimers.

\acknowledgments
The author thanks R.~Lazauskas  for discussions
and A.~C.~Fonseca for comments on the manuscript.


\begin{thebibliography}{10}
\expandafter\ifx\csname url\endcsname\relax\def\url#1{\texttt{#1}}\fi

\bibitem{efimov:plb}
\Name{Efimov V.} \REVIEW{Phys. Lett. B}{33}{1970}{563}.

\bibitem{braaten:rev}
\Name{Braaten E. \and Hammer H.-W.} \REVIEW{Phys. Rep.}{428}{2006}{259}.

\bibitem{kraemer:06a}
\Name{Kraemer T. {\it et al}} \REVIEW{Nature}{440}{2006}{315}.

\bibitem{knoop:09a}
\Name{Knoop S., Ferlaino F., Mark M., Berninger M., Sch\"obel H., N\"agerl
  H.-C. \and Grimm R.} \REVIEW{Nature Phys.}{5}{2009}{227}.

\bibitem{PhysRevLett.103.043201}
\Name{Barontini G., Weber C., Rabatti F., Catani J., Thalhammer G., Inguscio M.
  \and Minardi F.} \REVIEW{Phys. Rev. Lett.}{103}{2009}{043201}.

\bibitem{ferlaino:09a}
\Name{Ferlaino F., Knoop S., Berninger M., Harm W., D'Incao J.~P., N\"agerl
  H.-C. \and Grimm R.} \REVIEW{Phys. Rev. Lett.}{102}{2009}{140401}.

\bibitem{pollack:09a}
\Name{Pollack S.~E., Dries D. \and Hulet R.~G.}
  \REVIEW{Science}{326}{2009}{1683}.

\bibitem{deltuva:10c}
\Name{Deltuva A.} \REVIEW{Phys.~Rev.~A}{82}{2010}{040701(R)}.

\bibitem{hammer:07a}
\Name{Hammer H.~W. \and Platter L.} \REVIEW{Eur. Phys. J. A}{32}{2007}{113}.

\bibitem{stecher:09a}
\Name{von Stecher J., D'Incao J.~P. \and Greene C.~H.} \REVIEW{Nature
  Phys.}{5}{2009}{417}; see also Supplementary Information at
http://www.nature.com/nphys/journal/v5/n6/extref/nphys1253-s1.pdf

\bibitem{dincao:09a}
\Name{D'Incao J.~P., von Stecher J. \and Greene C.~H.} \REVIEW{Phys. Rev.
  Lett.}{103}{2009}{033004}.

\bibitem{deltuva:07a}
\Name{Deltuva A. \and Fonseca A.~C.} \REVIEW{Phys.~Rev.~C}{75}{2007}{014005}.

\bibitem{deltuva:07b}
\Name{Deltuva A. \and Fonseca A.~C.}
  \REVIEW{Phys.~Rev.~Lett.}{98}{2007}{162502}.

\bibitem{deltuva:07c}
\Name{Deltuva A. \and Fonseca A.~C.}
  \REVIEW{Phys.~Rev.~C}{76}{2007}{021001(R)}.

\bibitem{grassberger:67}
\Name{Grassberger P. \and Sandhas W.} \REVIEW{Nucl. Phys.B}{2}{1967}{181}.

\bibitem{alt:jinr}
\Name{Alt E.~O., Grassberger P. \and Sandhas W.} \REVIEW{JINR
  report}{E4-6688}{1972}{1}.

\bibitem{yakubovsky:67en}
\Name{Yakubovsky O.~A.} \REVIEW{Sov. J. Nucl. Phys.}{5}{1967}{937}.

\bibitem{deltuva:ef}
\Name{Deltuva A., Lazauskas R. \and Platter L.} \REVIEW{Few-Body
  Syst.}{}{2011}{to be published}.

\bibitem{deltuva:08a}
\Name{Deltuva A., Fonseca A.~C. \and Sauer P.~U.}
  \REVIEW{Phys.~Lett.~B}{660}{2008}{471}.

\bibitem{res_cpl}
\Name{Badalyan A.~M., Kok L.~P., Polikarpov M.~I. \and Simonov Y.~A.}
  \REVIEW{Phys. Rep.}{82}{1982}{31}.

\bibitem{lazauskas:he}
\Name{Lazauskas R. \and Carbonell J.} \REVIEW{Phys. Rev. A}{73}{2006}{062717}.

\end{thebibliography}

\end{document}